\documentclass[reprint,twocolumn,aps,pra,amsmath,amssymb,superscriptaddress,nofootinbib,longbibliography]{revtex4-2}
\usepackage{lipsum}  
\usepackage{ulem}
\usepackage{float}
\usepackage{layouts}
\usepackage{graphicx}
\usepackage{color}
\usepackage{xcolor}
\usepackage{mathtools}
\usepackage[bottom]{footmisc}
\usepackage{algorithmicx}
\usepackage{algpseudocode}
\usepackage[caption=false]{subfig}
\usepackage[percent]{overpic}

\begin{document}

\title{Simultaneous Intensity and Frequency Control for Optical Waveform Generation using an Acousto optic Modulator} 
		
\author{H. Madathil}
\affiliation{Center for Quantum Information and Control, Department of Physics and Astronomy, University of New Mexico, Albuquerque, New Mexico 87131}

\author{S. Wang}
\affiliation{Center for Quantum Information and Control, Department of Physics and Astronomy, University of New Mexico, Albuquerque, New Mexico 87131}

\author{F. E. Becerra}
\affiliation{Center for Quantum Information and Control, Department of Physics and Astronomy, University of New Mexico, Albuquerque, New Mexico 87131}
\email{fbecerra@unm.edu}

\begin{abstract}

We describe a method for calibrating the response of an acousto-optic modulator (AOM) to enable precise, arbitrary control of the intensity and frequency of optical fields. The method involves characterizing the nonlinear response of the AOM to its input RF drive voltage and applying an iterative calibration/correction algorithm to accurately map input RF amplitude and frequency drive to output optical intensity and frequency shifts. After a few calibration/correction iterations, 
the calibration of the AOM maintains the optical power within 1$\%$ of a constant target value over relatively wide frequency tuning range ($\approx$100 MHz), with $\sigma=0.2\%$ relative deviation. We apply this calibration method to generate programmable waveforms and optical pulses with tailored intensity and frequency profiles that closely match their target specifications. This technique offers a simple and robust method for applications that require high-accuracy optical modulation.
\end{abstract}

\maketitle
\section{Introduction}
 Precise control of optical fields is essential for diverse applications ranging from optical metrology and atomic spectroscopy \cite{hall1984external,letargat2006accurate} to quantum computing \cite{schaffer2018fast} and optical communications \cite{mesleh2018,lin2025}. Linear and nonlinear processes in crystals, such as acousto and electro-optic modulation, provide a way to enable power, frequency, phase, and spatial control of optical beams at relatively high bandwidths \cite{alexander2006photon_echoes,melnichuk2010direct_kerr,zhang2023plasmonic_metafibers,wongcampos2017two_atom_entanglement,jabbariFastControlTransverse2026}. Among different modulation techniques and devices for optical control, acousto-optic modulators (AOMs) are widely used across diverse experimental settings for controlling different degrees of freedom of light, including spatial \cite{braverman2020fast_generation}, temporal \cite{yang2025a}, and frequency \cite{negnevitsky2013wideband_locking}. AOMs provide a simple, cost-effective solution for high bandwidth and precise modulation with high optical damage thresholds. High-speed spatial modulation and single-pixel imaging \cite{liu2023aoslm}, and large frequency scanning employing multi-pass AOM configurations \cite{donley2005,zhou2020,decarlos2012triplepass},  are some examples of the extensive capabilities of AOM for controlling optical fields. However, achieving accurate arbitrary modulation often requires a precise calibration of the inherent nonlinear response of the AOM to the input radio frequency (RF) drive voltages. Moreover, while frequency scans can be achieved by varying the frequency of the RF input, the inherent nonlinearity of the AOMs produces large distortions in the output light intensity that are highly dependent on any frequency shifts. These nonlinear effects limit the applicability of AOMs for applications requiring simultaneous control of intensity and frequency.
 
We present a simple calibration method for characterizing and compensating for the nonlinear response of an AOM, enabling arbitrary control of optical intensity and frequency simultaneously. The method consists of measuring the frequency- and amplitude-dependent diffraction efficiency of the AOM and using this response to construct an iterative compensation algorithm. This allows the applied RF amplitude to be corrected during frequency modulation such that the optical intensity follows a desired target waveform with significantly reduced distortion. This approach turns the AOM from a device with a highly nonlinear response into a programmable actuator for optical-field frequency and intensity synthesis. In particular, it enables frequency sweeps with nearly constant optical intensity, and allows for user-defined waveforms in which the optical power and frequency can be varied independently with high accuracy. This capability is important in experiments where uncontrolled intensity variations during frequency modulation lead to systematic errors, reduced signal-to-noise ratio, or imperfect state preparation. Compared to approaches that rely on manual tuning or calibration at only a few operating points, our technique provides a systematic and experimentally straightforward route to high-fidelity waveform generation over a relatively wide range of RF amplitudes and frequencies. 

Because the procedure requires only standard AOM control hardware and direct optical power measurements, it can be readily implemented in many existing optical setups. The method therefore provides a practical advance for experiments requiring reproducible, high-bandwidth programmable control of optical fields. We demonstrate this capability by engineering optical waveforms with tailored intensity and frequency profiles, including compensated frequency sweeps and arbitrary modulation sequences.

\section{Background} \label{Sec-Background}

\begin{figure}[h]
    \centering
    \includegraphics[width=0.9\linewidth]{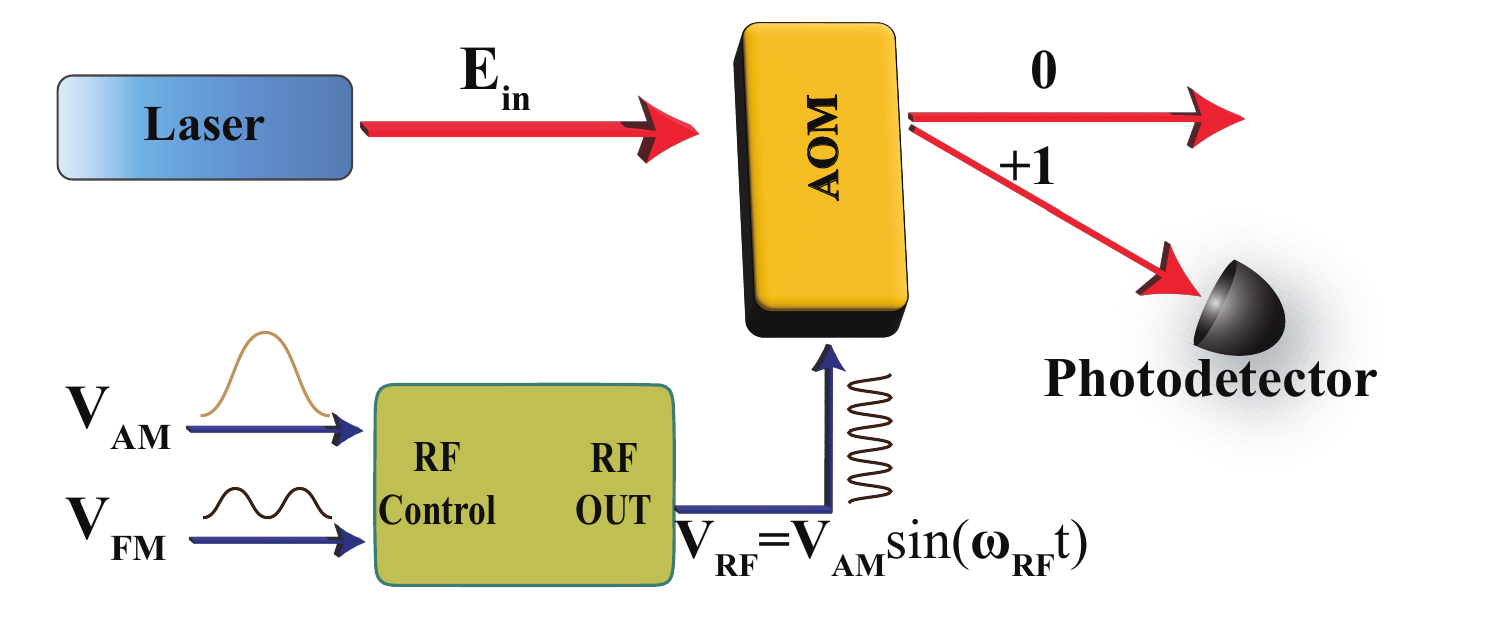}
    \caption{Optical field manipulation using an AOM. Two control voltages  $V_{AM}$ (amplitude) and $V_{FM}$ (frequency), generate an RF voltage  $V_{RF}$  with controllable amplitude and frequency. The AOM generates diffracted beams from an input field $E_{in}$ with frequency $\omega_0$. The field $E_{out}$ in the first diffraction order has a frequency and amplitude that depend on the control voltages $V_{AM}$ and $V_{FM}$, and $\omega_0$. }
    \label{SimpleSketch}
\end{figure}

\begin{figure*}[t]
    \centering
    \includegraphics[width=0.92\linewidth]{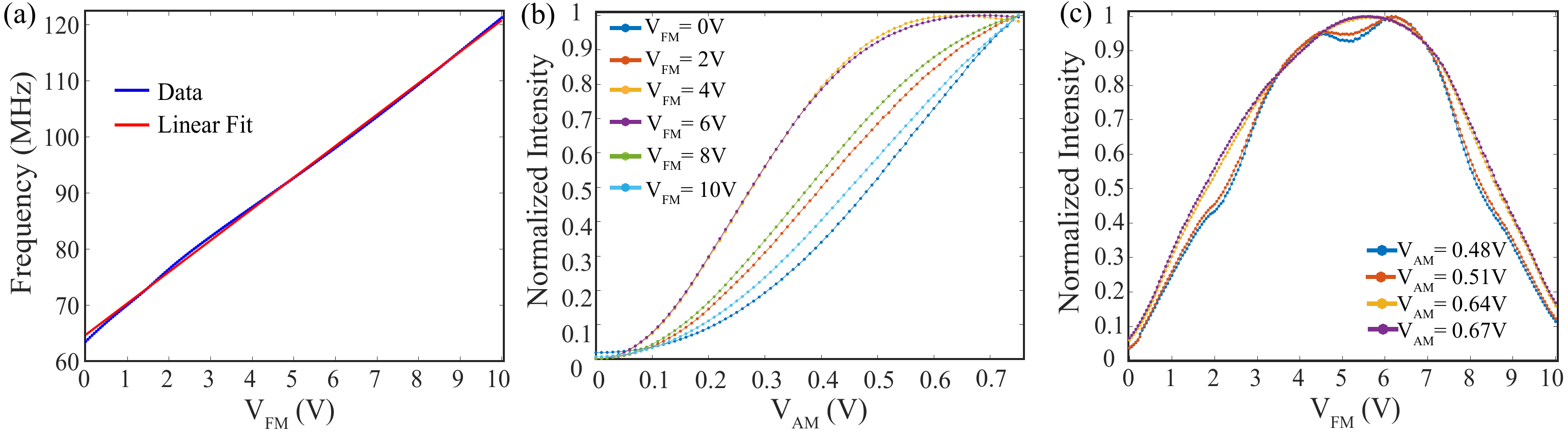}
    \caption{Frequency and amplitude response for the single-pass AOM. (a) (Linear) relationship between the control voltage $V_{FM}$  and frequency  $\omega_{RF}$ of $V_{RF}$, which is used to drive the AOM. (b) Diffraction efficiency curves into the $1^{st}$ diffraction order, as observed with a photo detector, as a function of $V_{AM}$ for several values of $V_{FM}$. (c) Diffraction efficiency curves for the $1^{st}$ diffraction order as a function of frequency, $V_{FM}$, for different $V_{AM}$ voltages.} 
    \label{AOMnlResponse}
\end{figure*}

AOMs are general purpose devices with a broad range of applications that require intensity, time, and frequency control of optical fields. This control is achieved by modulating the amplitude and frequency of the RF input voltage of the AOM. Figure \ref{SimpleSketch} shows the general setting, where an AOM is used to modulate the amplitude and frequency of an input optical field $E_{in}$ with a central frequency $\omega_0$. The AOM driven by an RF input voltage with a frequency $\omega_{RF}$ produces a dynamic diffraction grating, generating multiple beams from different diffraction orders.  In particular, the beam $E_{out}$ from the first diffraction order has a frequency $\omega_{out}=\omega_0+\omega_{RF}$, and, once properly aligned, an amplitude that depends on the amplitude of the RF input voltage $V_{RF}$ and the response of the AOM to the input $V_{RF}$. This dependence can be used to generate optical fields with tailored amplitude and frequency profiles over 100-400 MHz frequency spans.  However, while the amplitude and frequency of the $V_{RF}$ can be accurately controlled by control input voltages $V_{AM}$ and $V_{FM}$, respectively, the AOM has a nonlinear response to these control voltage(s) (see Fig. \ref{AOMnlResponse}). This nonlinearity makes it difficult, and non straightforward, to achieve precise simultaneous frequency and amplitude control of the output optical field $E_{out}$.

Figure \ref{AOMnlResponse} shows the response of a typical AOM to an input RF voltage $V_{RF}\approx V_{AM}\sin{\omega_{RF}t}$, with $\omega_{RF}\approx V_{FM}$. Figure \ref{AOMnlResponse}(a) shows the (linear) relationship between the control voltage $V_{FM}$ and the corresponding frequency $\omega_{RF}$ of $V_{RF}$. Because of energy and momentum conservation in the acousto-optic effect in AOMs, the frequency of the optical field in the first diffraction order (+1)  $E_{out}$ is $\omega_0+\omega_{RF}$. As a result, the linear dependence of  $\omega_{RF}$ with $V_{FM}$ allows for  directly controlling the frequency of  $E_{out}$, with a simple linear calibration of the RF source.  Here, we develop a simple method that takes into account, and corrects for, the  non-linear response of AOMs to RF input voltages that enables the generation of fields with tailored amplitude, temporal, and frequency profiles with high accuracy within relatively large frequency spans ( $>$ 100 MHz), and with very small intensity variations ($<1\%$).  

\section{Method}\label{Sec-Method}

The main idea of the proposed method is to use the information about the nonlinear intensity response of the AOM during a frequency sweep over a specified range of frequencies (See Fig. \ref{AOMnlResponse}.(c)) to  obtain the required amplitude controlled voltages ($V_{AM}$) to correct for this non-linear response. Then, apply iterative corrections to the intensity during the specified frequency sweep. After a few iterations, this method can correct for the intensity variations within $1\%$ over the desired frequency range. We describe this method in three steps.

\begin{figure*}
    \centering
    \includegraphics[width=0.92\linewidth]{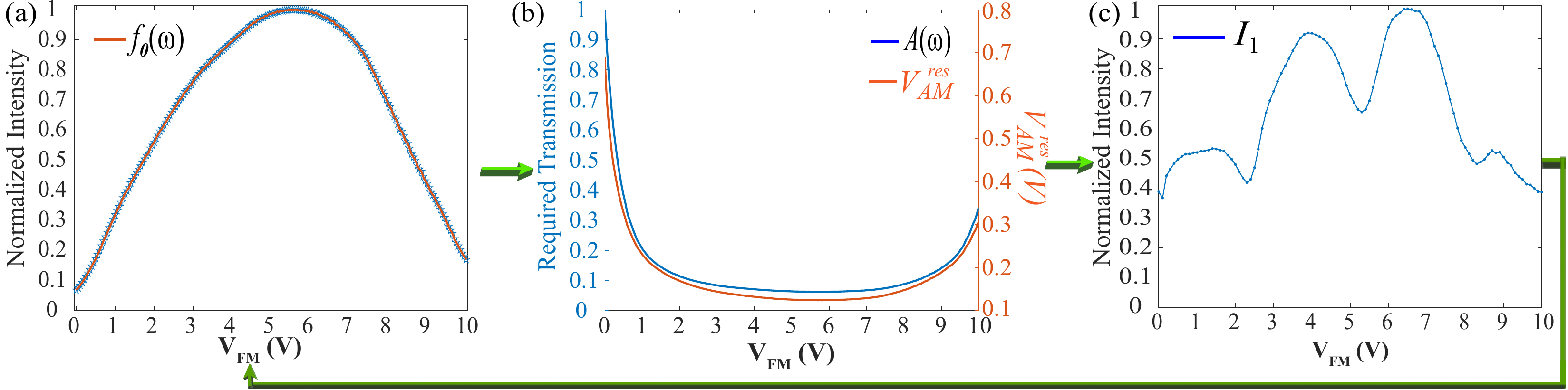}
    \caption{A schematic diagram that describes the iterative process associated with the intensity procedure. (a) Shows the initial frequency sweep of the AOM by varying the $V_{FM}$ voltages. (b) The reciprocal function needed for intensity calibration (blue) and the corresponding $V_{AM}^{res}$ required voltages (orange). (c) Frequency sweep after the first iteration making use of the reciprocal $V_{AM}^{res}$ voltages found in (b). }
    \label{model}
\end{figure*}

\textbf{1. }As a first step, the proposed method obtains a ``representative'' sample of the dependence of the AOM diffraction efficiency with respect to the amplitude control voltage $V_{AM}$. For example, this representative sample can be the average of the curves for different  $V_{FM}$ shown in Fig. \ref{AOMnlResponse}(b). Then, we model the dependence of the optical intensity $I$ of the AOM first-order diffraction mode as
\begin{equation}
    I=G(V_{AM})
    \label{I_G(Vam)}
\end{equation}
where $G$ is the corresponding nonlinear output response function over the full range of $V_{AM}$. Then we (numerically) obtain the inverse function of $G$ using the representative sample, denoted as $g=G^{-1}$ , which provides the control voltage $V_{AM}$ for a target intensity $I$: 
\begin{equation}
    V_{AM}=g(I)=G^{-1}(I)
    \label{Vam_g(I)}
\end{equation}
This inverse function $g$ will be well behaved as long as the function $G$ is bijective. In order to ensure this, the representative sample curve of $I$ vs. $V_{AM}$ should not contain regions with small slopes. \footnote{For our experimental implementation of the method, we used the gain curve for $V_{FM}=8$V, which is close to the average of the curves in Fig. \ref{AOMnlResponse}(b). We observe that this curve yields a very good calibration of the AOM nonlinearity.}

\textbf{2. } In the second step, we obtain the response curve for the output of the AOM as a function of frequency $V_{FM}$, as shown in Fig. \ref{AOMnlResponse}(c). This curve is then used to calibrate the response as a function of frequency, which is used to compensate for differences in diffraction efficiency for different frequencies. For this step, we set the amplitude modulation voltage $V_{AM}$ that corresponds to 90\% of the total normalized intensity  of the gain curve we selected in the previous step. This choice of $V_{AM}$ avoids regions of the curve with small slope, and provides robustness against drops of overall laser intensity. 

Figure \ref{model}(a) shows the normalized output intensity response for this particular choice of $V_{AM}$,  as a function of frequency control voltage $V_{FM}$. We denote this response function as $f{_{0}}(\omega)$. Next, we obtain the normalized reciprocal of the response function $f_{0}(\omega)$ for the frequency sweep:
\begin{equation}\label{eq:recipIntResp}
    A(\omega)=\frac{C_0}{f {_0}(\omega)}
\end{equation}
where $C_0=\min {f_0(\omega)}$ is a normalization constant, equal to the minimum of the response function $f_0(\omega)$ over the frequency sweep in Fig. \ref{model}(a).

\textbf{3.} The reciprocal function $A(\omega)$ (shown in Fig. \ref{model}(b)),
can be used to calculate the resultant amplitude modulation voltage $V_{AM}^{res}$, which is needed for correcting the nonlinear response of the AOM at different frequencies. These required voltages  $V_{AM}^{res}(\omega)$ for a given reciprocal transmission  $A(\omega)$ are obtained using Eq. (\ref{Vam_g(I)}), as
\begin{equation}
   V_{AM}^{res}(\omega)=g(A(\omega))\label{Vamres}
\end{equation}
Figure~\ref{model}(b) (red curve) shows $V_{AM}^{\mathrm{res}}(\omega)$ as a function of $V_{FM}$ for the corresponding reciprocal transmission  $A(\omega)$ (blue curve). The vector $V_{AM}^{\mathrm{res}}(\omega)$ is a function of $\omega$ which, in principle, can be used to correct the nonlinear response of the AOM with respect to both amplitude and frequency modulation.  As a result, applying an amplitude control voltage $V_{AM} = V_{AM}^{\mathrm{res}}(\omega)$ during a frequency sweep corresponding to the full range of $V_{FM}$, ideally would result in a constant output intensity across the entire frequency range. However, due to the intrinsic nonlinear response of the AOM, and that $V_{AM}^{\mathrm{res}}(\omega)$ was obtained from a single representative efficiency curve in Fig.~2, it is not guaranteed that a single iteration of this correction procedure would be sufficient for perfect compensation. In practice, we find that by applying this method iteratively, each iteration significantly reduces nonlinear effects in the AOM response, and the iterative method converges rapidly, requiring only a few iterations for near perfect compensation. 

Figure \ref{model} (c) shows the intensity of the first diffraction order as a function of frequency with an amplitude control voltage $V_{AM}=V_{AM}^{res}(\omega)$, showing a significantly smaller intensity variation over the full frequency span, as compared to Figure \ref{model}(a). This new intensity response function $I_{1}$ is then multiplied by the input response function $f_0(\omega)$ in that iteration, shown in Figure \ref{model} (a), to obtain the total response function up to that point $f_{\mathrm{1}}(\omega)\equiv f_0(\omega)*I_{1} $. Function $f_{\mathrm{1}}(\omega )$ can then be used as a new input response function in Eq.(\ref{eq:recipIntResp}) in a second iteration of this algorithm. As described in the following section, after a few iterations of this simple algorithm, it is straightforward to achieve relative intensity fluctuations well within 1\% for a wide range of frequencies. Moreover, by correcting for the nonlinear response of the AOM, this method enables the generation of programmable waveforms and optical pulses with tailored intensity and frequency profiles that closely match their target parameters. Appendix A contains a pseudo-code for the algorithm.

\section{Experiment}
\begin{figure}[h]
    \centering
    \includegraphics[width=0.85\linewidth]{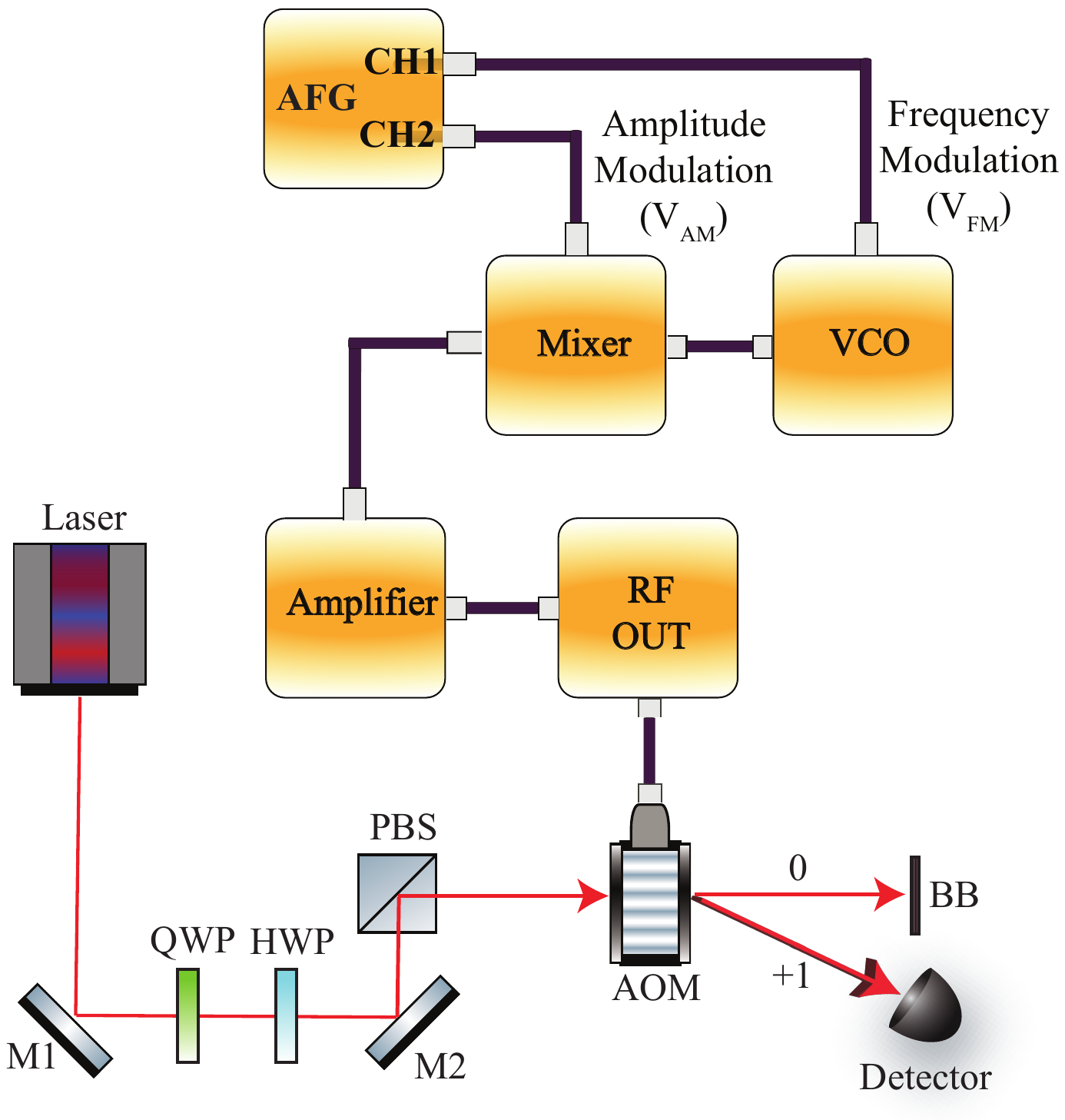}
    \caption{Experimental setup for intensity and frequency control based on an AOM.   
    HWP: Half-Wave Plate, M: Mirror, PBS: Polarizing Beam Splitter, AOM: Acousto-Optic Modulator, BB: Beam Blocker. } 
    \label{exp-setup}
\end{figure}

Figure \ref{exp-setup} shows the diagram of the experimental setup. We use an external cavity diode laser from MOGLABS at a wavelength of 852 nm. This laser with linear polarization is incident on an AOM from IntraAction Corp (ATM-1001A2), which operates at a central frequency of 100 MHz. We observe maximum diffraction efficiency in the first-order (+1) diffracted mode of 67\% compared to the total input power, when the frequency and amplitude of the input $V_{RF}$ to the AOM is optimized. This first-order diffraction mode is the output beam used  throughout the experiment. 

The $V_{RF}$ is generated from a voltage-controlled oscillator (VCO) from Mini-Circuits (ZOS-150), followed by an RF amplifier. The Frequency control of $V_{RF}$ is realized by  an input control voltage $V_{FM}$, generated from an arbitrary function generator (AFG), to the VCO, which allows for controlling the frequency from 60-150 MHz. Amplitude control of $V_{RF}$  is implemented using an RF mixer (Mini-Circuits, ZAD-1). By mixing an amplitude control voltage $V_{AM}$ with the output of the VCO, this mixer acts as a voltage-controlled attenuator (VCA). The resulting attenuation of  $V_{RF}$  shows a linear dependence with $V_{AM}$, which is generated by the AFG.
The output from the RF mixer is fed into the RF amplifier, which amplifies $V_{RF}$ to the required operating range for the AOM. We use a optical power meter (Thorlabs, PM100D) to implement the method described in Sec. \ref{Sec-Method} for the calibration and correction of the AOM nonlinearity. Subsequently, we use a switchable gain, fast photodetector (Thorlabs PDA100A2) to characterize the RF-engineered, time-dependent optical waveforms.   

During the calibration and correction stage, we use the AFG to remotely set the values for the control voltages $V_{FM}$ and $V_{AM}$ (see Figure \ref{exp-setup}) at rate of 2 Hz. The power meter then records the power of $E_{out}$ from the AOM at each voltage setting. For the generation of arbitrary waveforms, after the AOM calibration, the vectors of the settings for the control voltages $V_{FM}$ and $V_{AM}$ corresponding to the target frequency and amplitude waveforms are stored in memory of the AFG. Conditioned on an external trigger from an FPGA, these waveforms are sent to the VCO and the mixer to generate the desired optical waveform, which is measured by the fast photodetector (see Figure \ref{exp-setup}).

\section{Results}

\begin{figure*}[!htbp]
    \centering
    \includegraphics[width=0.95\linewidth]{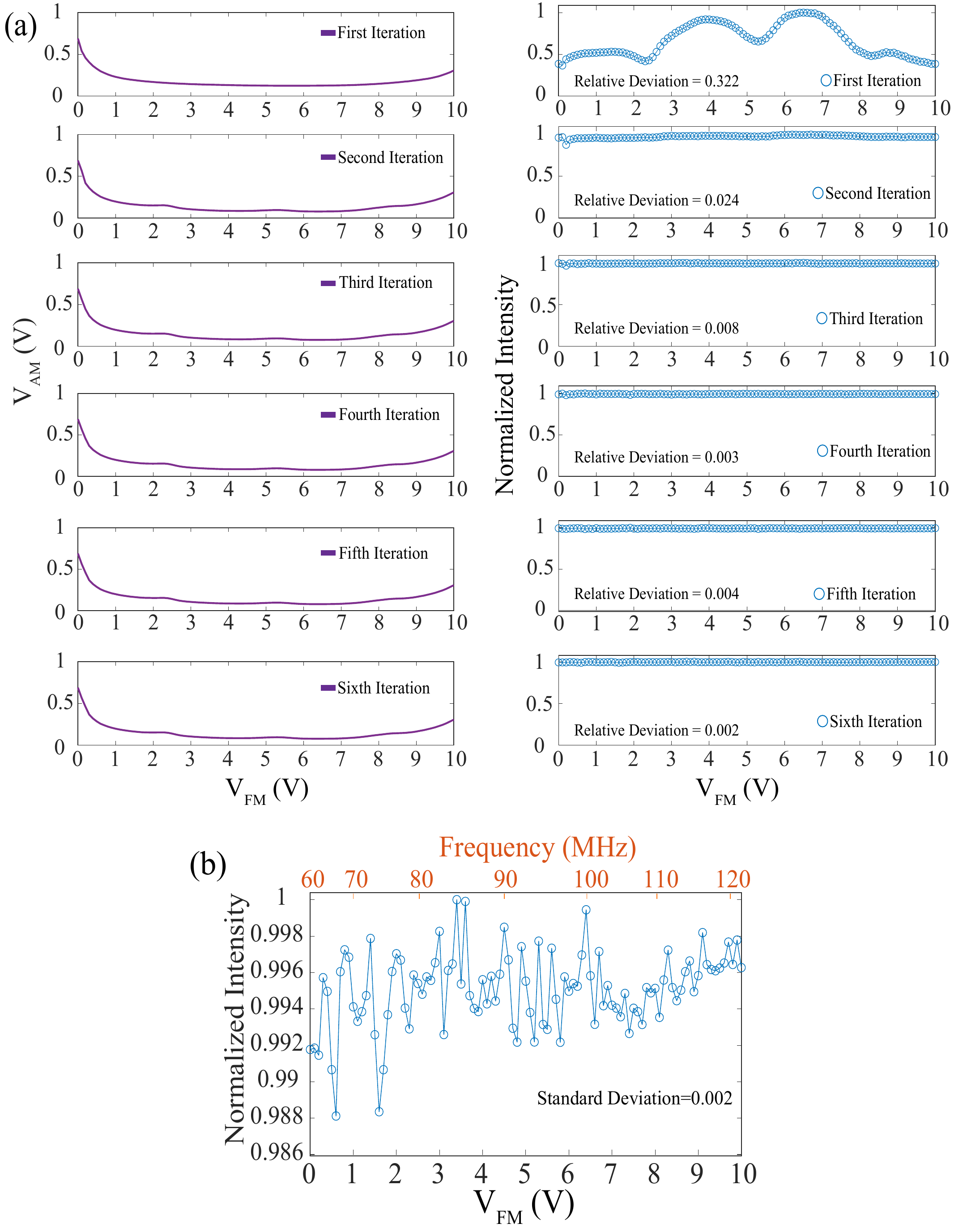}
    \caption{(a) Normalized intensity (left) and  corresponding amplitude control voltage $V_{AM}$  associated with a frequency sweep. It takes a total of six iteration to achieve a intensity variation within the 1\% range. (b) Zoom in of the normalized intensity after the (sixth) final iteration shown in (a).}
    \label{iter_result}
\end{figure*}







Figure~\ref{iter_result}.(a)  shows the results of the iterative compensation procedure. Initially, the first iteration follows the procedure illustrated in Figure \ref{model}. The initial intensity response $f_0(\omega)$ as a function of $V_{FM}$ in Fig. \ref{model}(a) is used to obtain the amplitude control voltage $V_{AM} = V_{AM}^{\mathrm{res}}(\omega)$ in Eq. (\ref{Vamres}) shown in Fig. \ref{model}(b), which is subsequently applied to correct for the AOM nonlinearity over this frequency interval in that iteration, as shown in Fig. \ref{model}(c).  The left panels in Figure~\ref{iter_result}.(a) show the amplitude-control voltages $V_{AM} = V_{AM}^{\mathrm{res}}(\omega)$ applied at each iteration. The right panels shows the corresponding output intensities for each iteration. We observe that successive iterations progressively make the output intensity closer to a constant value independently of $V_{FM}$. By the third iteration, the intensity variations are reduced to below 10\%, reaching approximately 1\% after the sixth iteration. Fig.~\ref{iter_result}.(b) shows a zoom in the normalized intensity after the sixth iteration, achieving an intensity within 1\% of the (normalized) target intensity, and with a  $\sigma$= 0.2\% relative deviation. We observed that additional iterations did not result in further measurable improvements. We attribute the observed stability limit to random fluctuations in the amplitude of the RF output voltage, which exhibits a relative standard deviation of approximately \(0.5\%\).
 
These results show that the proposed method for calibrating the  AOM non-linear response enables us to perform relatively wide frequency sweeps at nearly constant intensity. Constant-power frequency sweeps are a common tool for applications in atomic, molecular, and optical (AMO) physics, including absorption spectroscopy and nonlinear multi-wave mixing \cite{becerra2008,becerra2010,willis2009}. Beyond constant-intensity frequency sweeps, the same calibration method can enable accurate and fast simultaneous control of the amplitude and frequency of an optical field. This allows the AOM to be used as a programmable optical waveform generator rather than only as a frequency shifter or intensity modulator. In the following section, we demonstrate this capability by generating several representative optical waveforms with prescribed frequency and intensity profiles.

\begin{figure*}[!htbp]
    \centering
    \includegraphics[width=\linewidth]{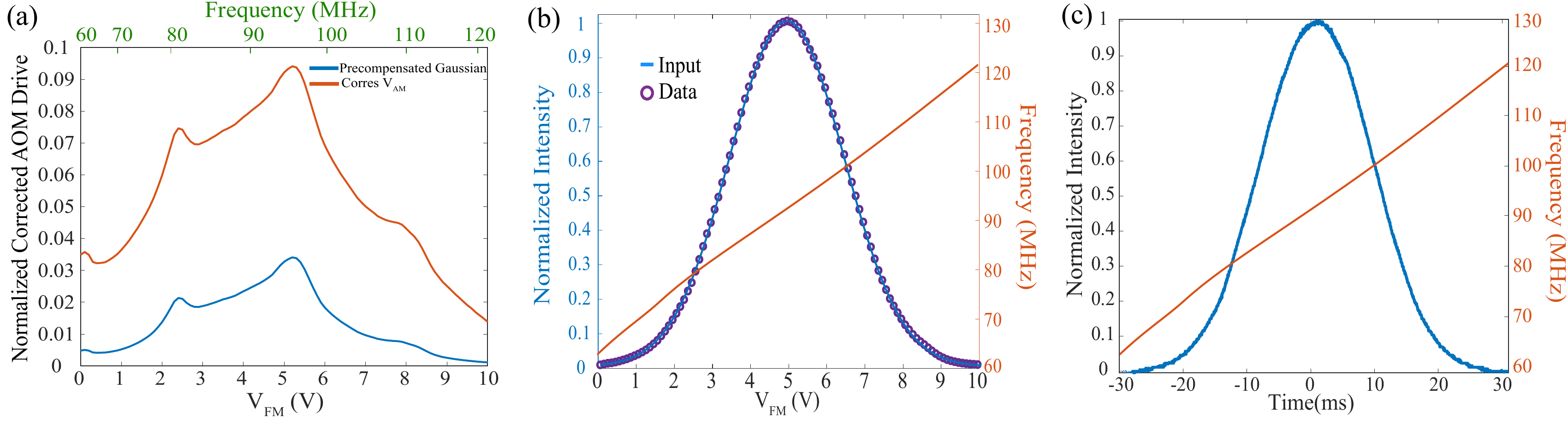}
    \caption{(a) Control voltage to generate a Gaussian intensity profile with a frequency sweep. (Blue) Pre-compensated Gaussian profile, which is the product of the normalized gaussian waveform with the reciprocal intensity response function for a constant-intensity sweep from 60 MHz to 120 MHz. (Orange) Amplitude control voltage $V_{AM}^{\mathrm{Gauss}}$ required to generate a Gaussian waveform with a frequency sweep. (b) Intensity of the AOM output $E_{out}$ as a function of  $V_{FM}$ (blue line and magenta circles), together with the measured VCO frequency (red). (c) Intensity of the Gaussian pulse $E_{out}$ as a function of time over 60 ms (blue), and measured frequency (red). }
    \label{linear_gauss}
\end{figure*}

\begin{figure*}[!htbp]
    \centering
    \includegraphics[width=\linewidth]{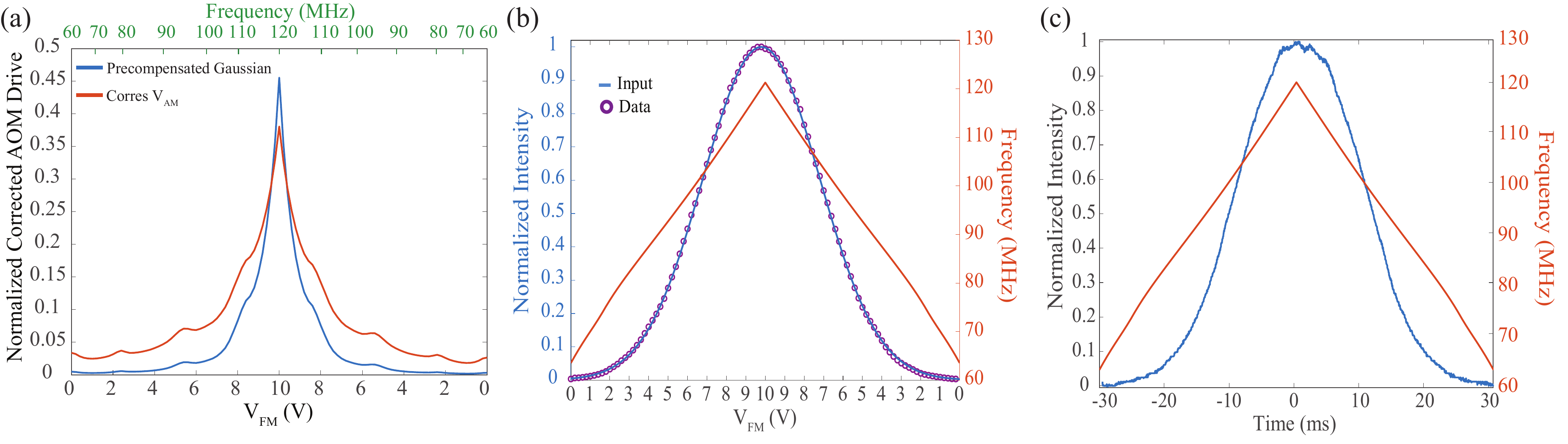}
    \caption{Generation of a Gaussian intensity waveform with a triangular frequency sweep. (a) Pre-compensated Gaussian waveform and corresponding control voltage $V_{AM}^{\mathrm{Gauss}}$ with a trangular frequency sweep from 60 MHz to 120 MHz. 
    (b) Intensity of the AOM output $E_{out}$ and measured VCO frequency as a function of  $V_{FM}$. (c) Generated Gaussian pulse $E_{out}(t)$  over 60 ms (blue), and measured frequency (red).}
    \label{triangle_sweep}
\end{figure*}

\subsection{Simple frequency Sweeps}

To demonstrate the flexibility of our method, we consider the generation of Gaussian intensity waveforms with two types of frequency sweeps: a linear sweep and a triangular sweep. To generate this waveform, we use the reciprocal intensity response function $A_{last}(\omega)$ obtained from the last iteration in the intensity-stabilization procedure in Fig.~\ref{iter_result}. Multiplying the target (Gaussian) waveform by $A_{last}(\omega)$ results in the pre-compensated Gaussian profile shown in blue in Fig.~\ref{linear_gauss}(a). Using the inverse function defined in Eq.~\ref{Vam_g(I)}, we then determine the amplitude control voltages $V_{AM}^{\mathrm{Gauss}}$ required to produce the target (Gaussian) waveform. We  prepare the Gaussian waveform with the linear frequency sweep using the AFG to generate both $V_{FM}$ and the corresponding $V_{AM}^{\mathrm{Gauss}}$ simultaneously. Fig. \ref{linear_gauss}(b) shows the normalized intensity of the AOM output $E_{out}$, measured with a power meter, as a function of  $V_{FM}$, together with the measured VCO frequency. Data points are taken at 1 Hz sampling rate. To generate a Gaussian pulse with a linear frequency sweep, we load the control voltage waveforms $V_{AM}^{\mathrm{Gauss}}$ and $V_{FM}$ in the AFG memory. Conditioned on an external trigger from an FPGA, these waveforms are sent/applied to the VCO and the mixer over 60 ms.  Fig.~\ref{linear_gauss}(c) shows the intensity of the generated Gaussian pulse $E_{out}(t)$ as a function of time, as measured by the fast photodiode, together with the frequency sweep from $\approx60-120$ MHz measured in the VCO.

    


As a second example, we prepared a Gaussian waveform with a triangular frequency sweep with a range from 62 MHz to 120 MHz. As in the case with linear-sweeps, we first determine the pre-compensated Gaussian intensity profile by the product of the target waveform and reciprocal intensity response function for the triangular frequency sweep. We then obtain the required amplitude control voltages $V_{AM}$ from Eq.~\ref{Vam_g(I)} to produce the Gaussian waveform with a triangular frequency sweep. 
Figure~\ref{triangle_sweep}(a) shows the pre-compensated Gaussian intensity profile and the corresponding $V_{AM}^{\mathrm{Gauss}}$ as a function of   $V_{FM}$. 
Fig.~\ref{triangle_sweep}(b) shows the intensity of $E_{out}$ as a function of  $V_{FM}$, together with the measured VCO frequency, acquired at a 2 Hz sampling rate in a powermeter. Fig.~\ref{triangle_sweep}(c) shows the generated Gaussian pulse $E_{out}(t)$ over 60 ms with a triangular frequency sweep. We observe that the intensity and frequency profiles for the generated pulse closely follow the target waveforms.



\subsection{Double-Pass AOM Configuration}

The frequency range for waveform generation using AOMs in a single pass configuration is limited by the bandwidth of the AOM, beyond which the diffraction efficiency reduces significantly. By contrast, multipass configurations allow for much wider frequency spans while preserving fast tunability \cite{donley2005,decarlos2012triplepass,zhou2020}.
The proposed AOM calibration method can be implemented in multi-pass configurations to generate optical programmable waveforms over much wider frequency ranges. As a proof of principle, we implement this method in a double-pass configuration, allowing for doubling the optical frequency span given a RF drive.

\begin{figure}[h]
    \centering
    \includegraphics[width=\linewidth]{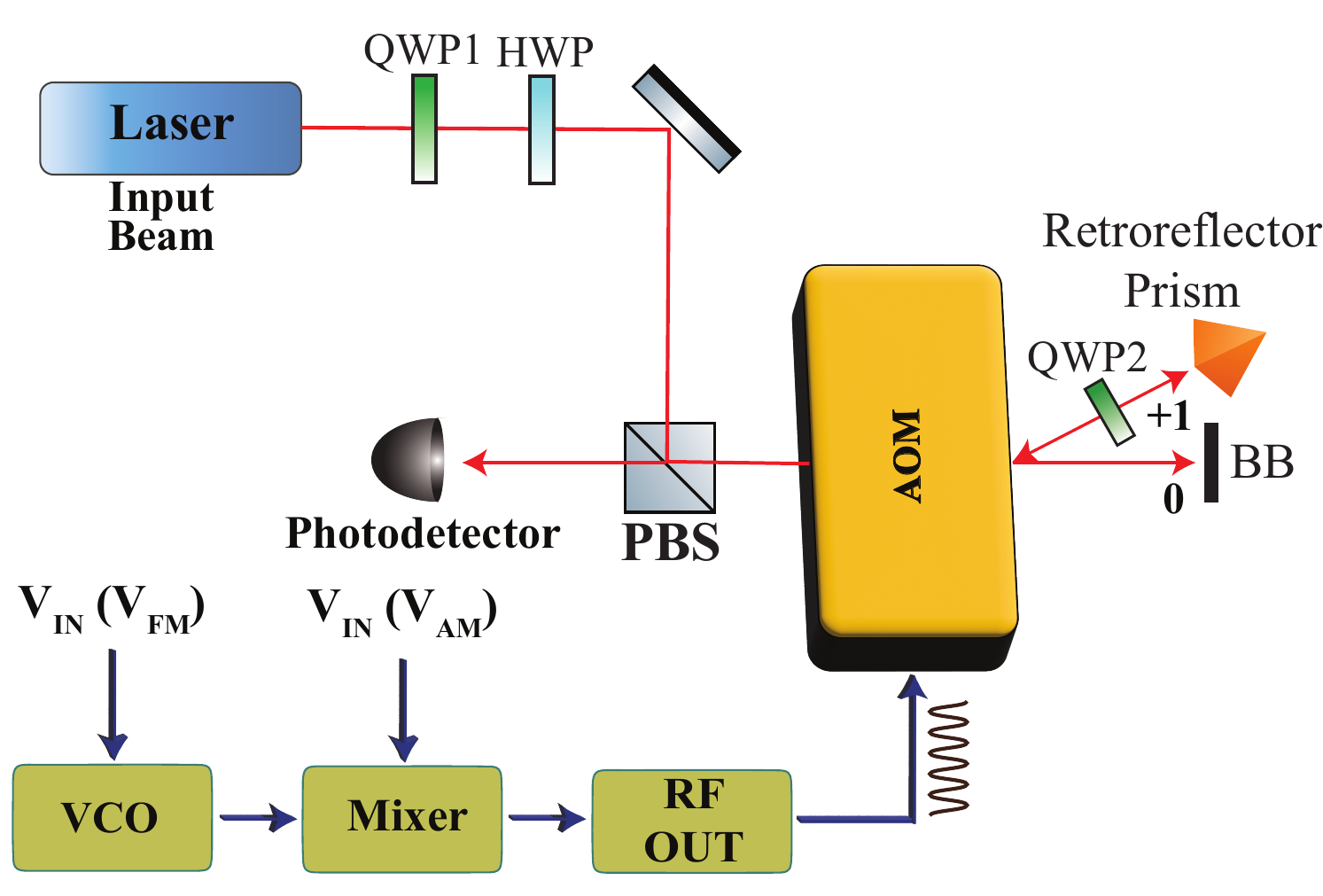}
    \caption{Experimental setup for the double-pass configuration. HWP: half-wave plate, QWP: quarter-wave plate, M: mirror, PBS: polarizing beam splitter, AOM: acousto-optic modulator, BB: beam blocker.}
    \label{1st_dp}
\end{figure}

 Figure~\ref{1st_dp} shows the experimental setup for the double pass configuration. The input beam passes through the AOM, and the first order diffracted beam is reflected back by a cat-eye retroreflector, passing through the AOM for a second time. The output beam after the second pass acquires a frequency shift equal to twice the RF input frequency $\omega_{RF}$, allowing for a substantially wider optical sweep range.

\begin{figure}[h]
    \centering
    \includegraphics[width=\linewidth]{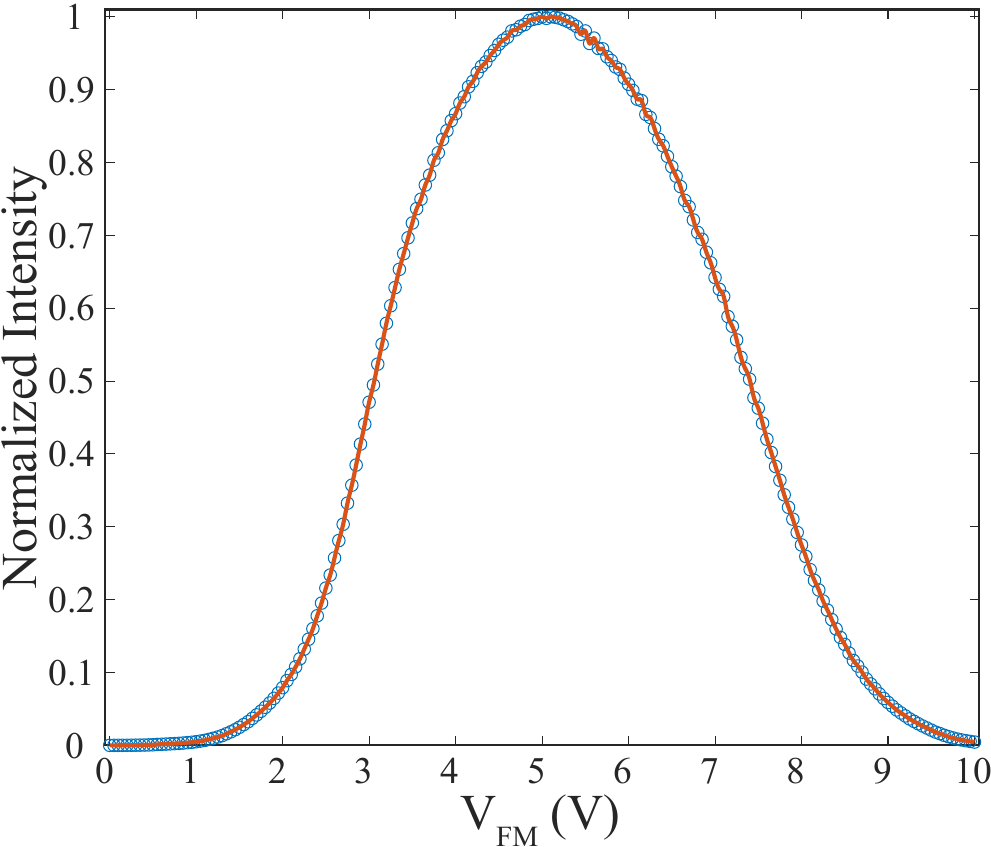}
    \caption{Normalized optical intensity during the initial uncompensated frequency sweep in the double-pass AOM configuration.}
    \label{zero_dp}
\end{figure}

\begin{figure}[h]
    \centering
    \includegraphics[width=\linewidth]{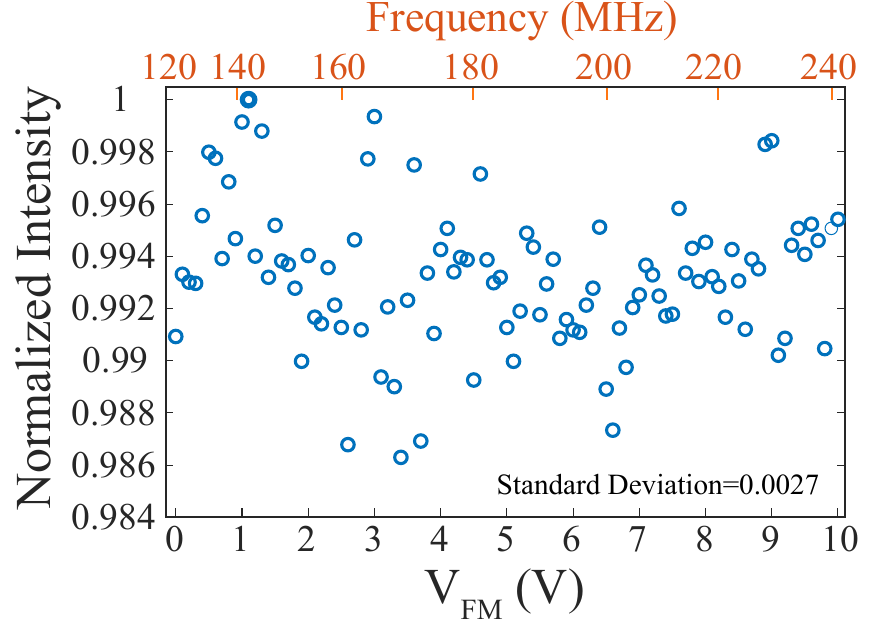}
    \caption{Zoomed-in view of the final corrected frequency sweep obtained after fourteen iterations of the compensation algorithm in the double-pass configuration. In this geometry, the optical frequency shift ($\approx100$ MHz) is twice the RF drive frequency shift. }
    \label{14th_dp}
\end{figure}

Figure~\ref{zero_dp}  shows the normalized optical intensity after the second pass during a frequency sweep without amplitude correction, showing the AOM response in the double pass configuration. Note that the intensity profile is substantially narrower than in the single pass configuration in Figure \ref{model}(a) due to the frequency-dependent diffraction efficiency per pass. While this effect makes compensation of the nonlinear response of the AOM more difficult, we find that our iterative method can nonetheless compensate for such nonlinearities. 
Figure~\ref{14th_dp} shows the result of the AOM calibration in the double pass configuration for obtaining a constant intensity over a frequency sweep, displayed for a range of 0.98-1. We observe that the iterative procedure converges more slowly compared to the single-pass configuration, as expected from the stronger effective nonlinearity of the two-pass response in Fig.~\ref{zero_dp}. Specifically, fourteen iterations were required to reach a performance level comparable to that obtained in the single-pass case with six iterations, achieving a relative standard deviation of $\sigma_{\mathrm{rel}}=2.7\times10^{-2}$ over $\approx100$ MHz as shown in Figure \ref{14th_dp}.  Moreover, this method can be used in multipass configurations, further widening the range of frequencies for which accurate intensity stabilization can be achieved.

\begin{figure}[h]
    \centering
    \includegraphics[width=\linewidth]{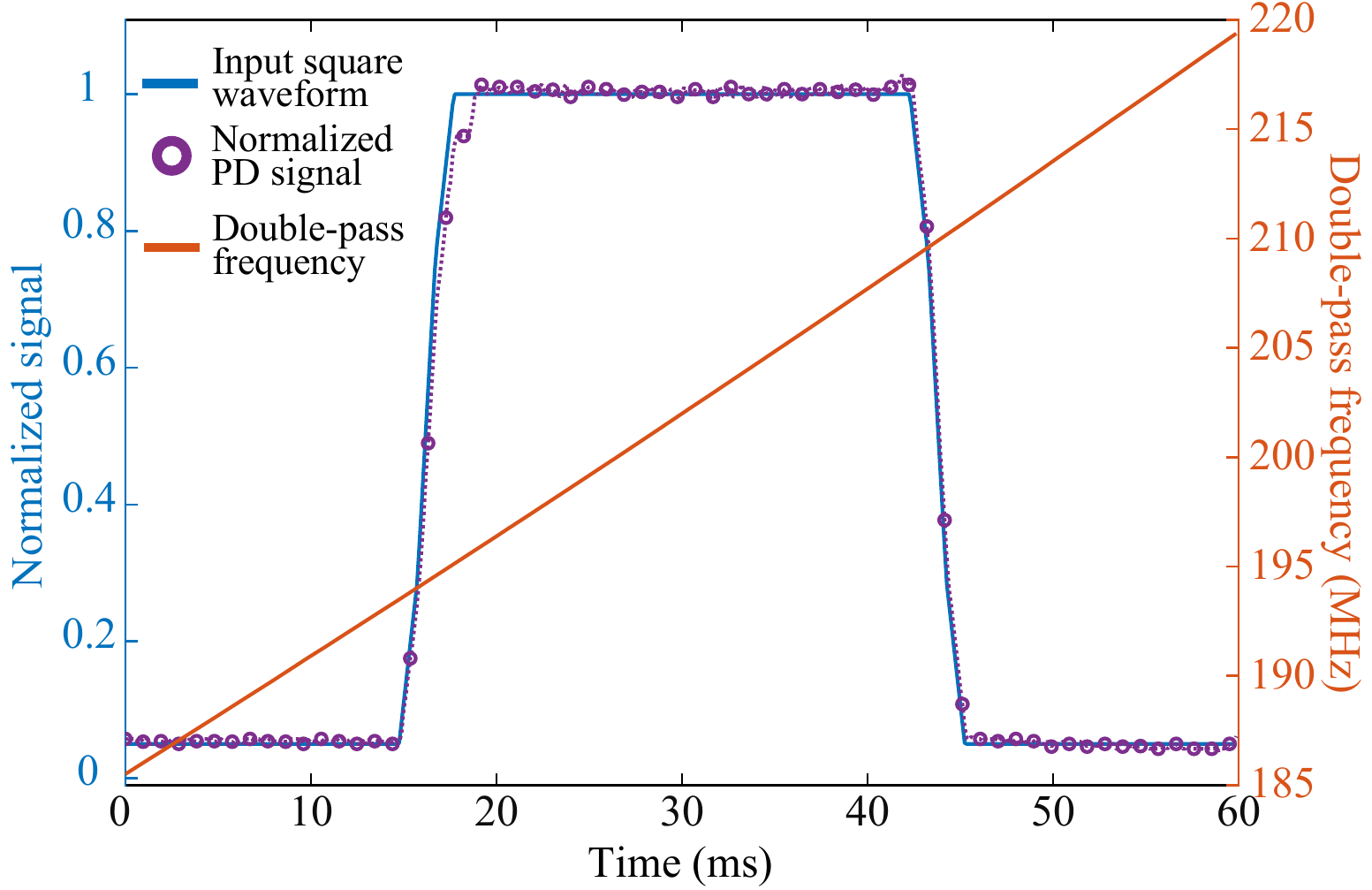}
    \caption{Example of waveform engineering in the double-pass configuration, showing a square intensity envelope combined with a frequency sweep over an extended optical frequency range. Such a waveform can be used to probe a broad atomic absorption profile in a single shot.}
    \label{square_dp}
\end{figure}

Figure \ref{square_dp} shows an example for the generation of a  flat-top optical pulse with a linear frequency sweep in the double pass configuration. We observe a good agreement with the target intensity and frequency profiles. These pulses can be used to probe broad atomic lines in a single shot over an extended frequency range. 

\subsection{Arbitrary Frequency Waveforms Beyond Simple Sweeps}

The method presented here enables the generation of arbitrary optical waveforms with independently programmable frequency and intensity profiles that closely reproduce the desired temporal profiles. 
Consider a target optical waveform with frequency $\omega_{target}(t)$ and intensity $I_{\mathrm{target}}(t)$ temporal profiles. From the calibration of the non-linear response of the AOM, we can determine the corresponding temporal correction waveform required for this frequency profile $\omega_{target}(t)$. Specifically, given the target $I_{\mathrm{target}}(t)$, the required pre-compensated profile $T_{\mathrm{req}}(t)$ is obtained by multiplying $I_{\mathrm{target}}(t)$ by the reciprocal correction function derived from the calibration in Eq. (\ref{eq:recipIntResp}) for $\omega_{target}(t)$:
\begin{equation}
T_{\mathrm{req}}(t) = I_{\mathrm{target}}(t)\,\frac{C_n}{f_n[\omega_{target}(t)]}.
\end{equation}
Using the inverse function defined in Eq.~(\ref{Vam_g(I)}), we obtain the required amplitude control  as a function of time: 
\begin{equation}\label{eq:V_AM_Arb}
V_{AM}(t)=g\!\left[T_{\mathrm{req}}(t)\right]
=
g\!\left[I_{\mathrm{target}}(t)\frac{C_n}{f_n[\omega_{target}(t)]}\right].
\end{equation}
Therefore, to generate waveform with specific  frequency $\omega_{target}(t)$ and intensity $I_{\mathrm{target}}(t)$ profiles, we use the frequency-control waveform $V_{FM}=\omega_{target}(t)$  and the amplitude-control waveform $V_{AM}(t)$ above to drive the AOM.  

\begin{figure}[h]
    \centering
    \includegraphics[width=\linewidth]{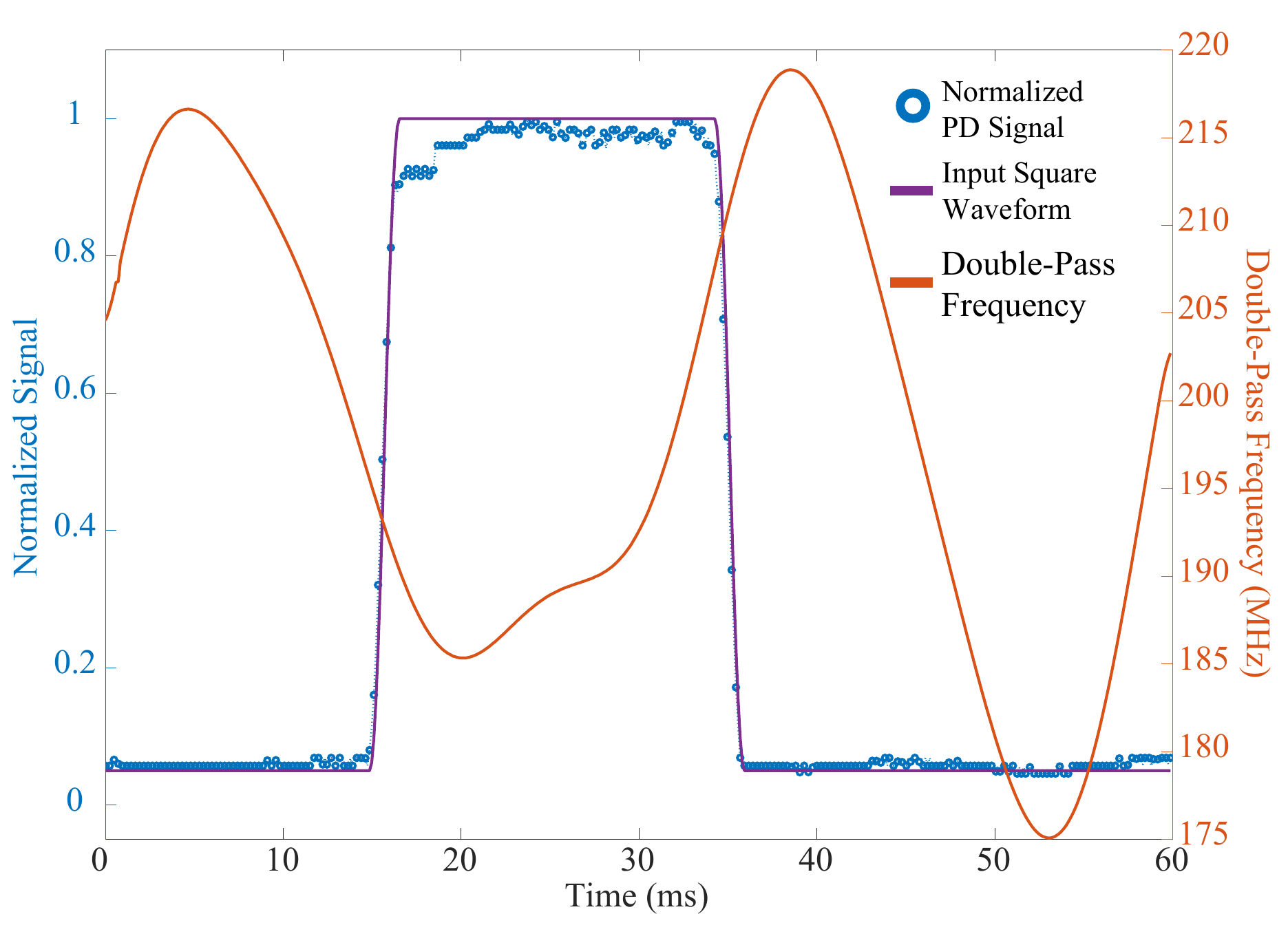}
    \caption{An example of waveform engineering in the double pass configuration, shwoing a square intesnity envelope combined with an arbitrary frequency sweep.}
    \label{arb_fig}
\end{figure}

Figure~\ref{arb_fig} shows an example of an optical waveform generated with complex temporal frequency and intensity profiles. This example demonstrates the versatility of the method, enabling an AOM to generate arbitrary optical waveforms with prescribed amplitude and frequency profiles with high accuracy. We observe that the complex frequency modulation introduces distortions in the temporal intensity profile of the waveform. We believe these distortions arise from transient effects in the RF drive circuitry, including parasitic capacitances in the RF signal path and the finite dynamic response of components such as the RF mixer operated as a voltage-controlled attenuator. Together, these effects can introduce phase shifts and transient variations in the generated RF signal, leading to the observed distortions. These effects can be mitigated through improved RF signal-generation hardware and optimized circuit design, for example by using integrated RF platforms such as RFSoC-based systems. As an alternative solution, an extension of the method described here could be applied to correct for these small transient effects iteratively to minimize deviations from the target waveform $I_{\mathrm{target}}(t)$ using Eq. (\ref{eq:V_AM_Arb}).


\section{Discussion and Conclusion}

We demonstrate a method that uses a simple iterative procedure to accurately calibrate the nonlinear response of an acousto-optic modulator for simultaneous control of optical intensity and frequency. The method suppresses the intensity distortions that ordinarily accompany frequency sweeps and enables the generation of user-defined optical waveforms with high accuracy. An important feature of the approach is that it does not require a detailed physical model of the AOM. Instead, it relies on direct experimental characterization of the device and iterative correction of the applied amplitude and frequency-modulation waveform.

Using the iterative compensation procedure, we achieved nearly constant optical intensity during frequency sweeps, with a relative standard deviation of approximately \(0.2\%\). We further demonstrated the applicability of the method in both single-pass and double-pass configurations. In each case, the calibrated response was used to suppress frequency-dependent intensity variations and to generate optical waveforms with prescribed amplitude and frequency profiles.

This method can be implemented using standard laboratory hardware, including voltage-controlled oscillators, voltage-controlled attenuators, arbitrary function generators, and optical power measurements. It can therefore be incorporated into existing optical setups with minimal overhead. The calibration procedure can also complement more advanced RF-control platforms, including Radio Frequency System-on-Chip (RFSoC) hardware. Such platforms enable rapid and highly programmable generation of RF amplitude and frequency waveforms, while the AOM calibration method developed here provides the required waveforms to compensate for the nonlinear optical response of the AOM to the RF input voltage. Combining these capabilities can therefore enable precise generation of optical waveforms in demanding applications such as robust quantum control and quantum information processing \cite{ruzic2024}.

\section*{Acknowledgments}
Work supported by the Department of Defense (DoD) Grant \# W911NF-23-1- 0246; the National Science Foundation (NSF Award No. MCB-2444171, PHY-2210447, PHY-2609629), and the Department of Energy (DOE) Contract No. CW42943.

\bibliography{references}    

@article{hall1984external,
  title = {External dye-laser frequency stabilizer},
  author = {Hall, J. L. and Hansch, T. W.},
  journal = {Optics Letters},
  volume = {9},
  number = {11},
  pages = {502--504},
  year = {1984},
  publisher = {Optical Society of America},
  doi = {10.1364/OL.9.000502}
}

@article{donley2005,
  title = {A Compact Double-Pass Acousto-Optic Modulator System},
  author = {Donley, Elizabeth A. and Heavner, Thomas P. and Levi, F. and Tataw, M. O. and Jefferts, Steven R.},
  year = 2005,
  month = jun,
  journal = {Review of Scientific Instruments},
  volume = {76},
  number = {063112},
  publisher = {Elizabeth A. Donley, Thomas P. Heavner, F Levi, M O. Tataw, Steven R. Jefferts},
  urldate = {2026-08-06},
 
}

@article{decarlos2012triplepass,
  title = {Laser frequency shifting by using two novel triple-pass acousto-optic modulator configurations},
  author = {de Carlos-López, E. and López, J. M. and López, S. and Espinosa, M. G. and Lizama, L. A.},
  journal = {Review of Scientific Instruments},
  volume = {83},
  number = {11},
  pages = {116102},
  year = {2012},
  doi = {10.1063/1.4758998}
}

@article{zhou2020,
  title = {Laser Frequency Shift up to 5 {{GHz}} with a High-Efficiency 12-Pass 350-{{MHz}} Acousto-Optic Modulator},
  author = {Zhou, Chao and He, Chuan and Yan, Si-Tong and Ji, Yu-Hang and Zhou, Lin and Wang, Jin and Zhan, Ming-Sheng},
  year = 2020,
  month = mar,
  journal = {Review of Scientific Instruments},
  volume = {91},
  number = {3},
  pages = {033201},
  issn = {0034-6748},
  doi = {10.1063/1.5142314},
  urldate = {2026-08-06},

}

@article{liu2023aoslm,
  title = {Using an acousto-optic modulator as a fast spatial light modulator},
  author = {Liu, Xialin and Braverman, Boris and Boyd, Robert W.},
  journal = {Optics Express},
  volume = {31},
  number = {2},
  pages = {1501--1515},
  year = {2023},
  doi = {10.1364/OE.31.001501}
}

@article{ruzic2024,
  title = {Leveraging Motional-Mode Balancing and Simply Parametrized Waveforms to Perform Frequency-Robust Entangling Gates},
  author = {Ruzic, Brandon P. and Chow, Matthew N.H. and Burch, Ashlyn D. and Lobser, Daniel S. and Revelle, Melissa C. and Wilson, Joshua M. and Yale, Christopher G. and Clark, Susan M.},
  year = 2024,
  month = jul,
  journal = {Physical Review Applied},
  volume = {22},
  number = {1},
  pages = {014007},
  publisher = {American Physical Society},
  doi = {10.1103/PhysRevApplied.22.014007},
  urldate = {2026-08-05},
  
}

@article{schaffer2018fast,
  title = {Fast quantum logic gates with trapped-ion qubits},
  author = {Schäfer, V. and Ballance, C. J. and Thirumalai, K. and Stephenson, L. J. and Ballance, T. G. and Steane, A. M. and Lucas, D. M.},
  journal = {Nature},
  volume = {555},
  pages = {75--78},
  year = {2018},
  doi = {10.1038/nature25737}
}

@article{mesleh2018,
  title = {Acousto-{{Optical Modulators}} for {{Free Space Optical Wireless Communication Systems}}},
  author = {Mesleh, Raed and {AL-Olaimat}, Ayat},
  year = 2018,
  month = may,
  journal = {Journal of Optical Communications and Networking},
  volume = {10},
  number = {5},
  pages = {515--522},
  publisher = {Optica Publishing Group},
  issn = {1943-0639},
  doi = {10.1364/JOCN.10.000515},
  urldate = {2026-08-06},
  
}

@article{letargat2006accurate,
  author       = {Le Targat, Rodolphe and Baillard, Xavier and Fouch{\'e}, Mathilde and Brusch, Anders and Tcherbakoff, Olivier and Rovera, Giovanni D. and Lemonde, Pierre},
  title        = {Accurate optical lattice clock with {Sr} atoms},
  journal      = {Physical Review Letters},
  volume       = {97},
  number       = {13},
  pages        = {130801},
  year         = {2006},
  month        = sep,
  day          = {26},
  doi          = {10.1103/PhysRevLett.97.130801}
}

@article{alexander2006photon_echoes,
  author    = {Alexander, A. L. and Longdell, J. J. and Sellars, M. J. and Manson, N. B.},
  title     = {Photon echoes produced by switching electric fields},
  journal   = {Physical Review Letters},
  volume    = {96},
  number    = {4},
  pages     = {043602},
  year      = {2006},
  month     = feb,
  day       = {1},
  doi       = {10.1103/PhysRevLett.96.043602}
}

@article{melnichuk2010direct_kerr,
  author    = {Melnichuk, Mike and Wood, Lowell T.},
  title     = {Direct Kerr electro‑optic effect in noncentrosymmetric materials},
  journal   = {Physical Review A},
  volume    = {82},
  number    = {1},
  pages     = {013821},
  year      = {2010},
  month     = jul,
  day       = {21},
  doi       = {10.1103/PhysRevA.82.013821}
}

@article{zhang2023plasmonic_metafibers,
  author       = {Zhang, Lei and Sun, Xinyu and Yu, Hongyan and Deng, Niping and Qiu, Feng and Wang, Jiyong and Qiu, Min},
  title        = {Plasmonic metafibers electro‑optic modulators},
  journal      = {Light: Science \& Applications},
  volume       = {12},
  number       = {1},
  pages        = {198},
  year         = {2023},
  month        = aug,
  day          = {22},
  doi          = {10.1038/s41377-023-01255-7}
}

@article{wongcampos2017two_atom_entanglement,
  author    = {Wong-Campos, Jaime D. and Moses, Steven A. and Johnson, Kevin G. and Monroe, Christopher},
  title     = {Demonstration of Two-Atom Entanglement with Ultrafast Optical Pulses},
  journal   = {Physical Review Letters},
  volume    = {119},
  number    = {23},
  pages     = {230501},
  year      = {2017},
  month     = dec,
  doi       = {10.1103/PhysRevLett.119.230501}
}

@article{braverman2020fast_generation,
  author    = {Braverman, Boris and Skerjanc, Alexander and Sullivan, Nicholas and Boyd, Robert W.},
  title     = {Fast generation and detection of spatial modes of light using an acousto‑optic modulator},
  journal   = {Optics Express},
  volume    = {28},
  number    = {20},
  pages     = {29112--29121},
  year      = {2020},
  month     = sep,
  day       = {28},
  doi       = {10.1364/OE.404309}
}

@article{negnevitsky2013wideband_locking,
  author    = {Negnevitsky, Vlad and Turner, Lincoln D.},
  title     = {Wideband laser locking to an atomic reference with modulation transfer spectroscopy},
  journal   = {Optics Express},
  volume    = {21},
  number    = {3},
  pages     = {3103--3113},
  year      = {2013},
  month     = mar,
  doi       = {10.1364/OE.21.003103},
  
}

@article{becerra2010,
  title = {Nondegenerate Four-Wave Mixing in Rubidium Vapor: {{Transient}} Regime},
  shorttitle = {Nondegenerate Four-Wave Mixing in Rubidium Vapor},
  author = {Becerra, F. E. and Willis, R. T. and Rolston, S. L. and Carmichael, H. J. and Orozco, L. A.},
  year = 2010,
  month = oct,
  journal = {Physical Review A},
  volume = {82},
  number = {4},
  pages = {043833},
  publisher = {American Physical Society},
  doi = {10.1103/PhysRevA.82.043833},
  urldate = {2026-07-16}
}

@article{becerra2008,
  title = {Nondegenerate Four-Wave Mixing in Rubidium Vapor: {{The}} Diamond Configuration},
  shorttitle = {Nondegenerate Four-Wave Mixing in Rubidium Vapor},
  author = {Becerra, F. E. and Willis, R. T. and Rolston, S. L. and Orozco, L. A.},
  year = 2008,
  month = jul,
  journal = {Physical Review A},
  volume = {78},
  number = {1},
  pages = {013834},
  issn = {1050-2947, 1094-1622},
  doi = {10.1103/PhysRevA.78.013834},
  urldate = {2026-07-15},
  copyright = {http://link.aps.org/licenses/aps-default-license},
  langid = {english}
}

@article{willis2009,
  title = {Four-Wave Mixing in the Diamond Configuration in an Atomic Vapor},
  author = {Willis, R. T. and Becerra, F. E. and Orozco, L. A. and Rolston, S. L.},
  year = 2009,
  month = mar,
  journal = {Physical Review A},
  volume = {79},
  number = {3},
  pages = {033814},
  publisher = {American Physical Society},
  doi = {10.1103/PhysRevA.79.033814},
  urldate = {2026-07-16}
}

@article{jabbariFastControlTransverse2026,
  title = {Fast Control of the Transverse Structure of a Light Beam Using Acousto-Optic Modulators},
  author = {Jabbari, Mahdieh Chartab and Li, Cheng and Liu, Xialin and {C{\'o}rdova-Castro}, R. Margoth and Braverman, Boris and Upham, Jeremy and Boyd, Robert W.},
  year = 2026,
  month = may,
  journal = {Physical Review Applied},
  volume = {25},
  number = {5},
  pages = {054055},
  issn = {2331-7019},
  doi = {10.1103/91kv-cxlp},
  urldate = {2026-06-18},
  langid = {english},
 
}

@article{lin2025,
  title = {Optical Multi-Beam Steering and Communication Using Integrated Acousto-Optics Arrays},
  author = {Lin, Qixuan and Fang, Shucheng and Yu, Yue and Xi, Zichen and Shao, Linbo and Li, Bingzhao and Li, Mo},
  year = 2025,
  month = may,
  journal = {Nature Communications},
  volume = {16},
  number = {1},
  pages = {4501},
  publisher = {Nature Publishing Group},
  issn = {2041-1723},
  doi = {10.1038/s41467-025-59831-x},
  urldate = {2026-08-06},
  copyright = {2025 The Author(s)},
  langid = {english}
}

@article{yang2025a,
  title = {Compact Arbitrary Optical Waveform Modulator with Digital Feedback},
  author = {Yang, Shuzhe and Masella, Guido and Moeini, Vase and Bellahsene, Amar and Li, Chang and Bienaim{\'e}, Tom and Whitlock, Shannon},
  year = 2025,
  month = may,
  journal = {Physical Review Applied},
  volume = {23},
  number = {5},
  pages = {054009},
  issn = {2331-7019},
  doi = {10.1103/PhysRevApplied.23.054009},
  urldate = {2026-08-06},
  langid = {english}
}



\newlength{\appendixcolumnwidth}
\setlength{\appendixcolumnwidth}{\columnwidth}

\onecolumngrid
\appendix

\noindent
\begin{minipage}[t]{\appendixcolumnwidth}
\vspace{0.5pt}
\section{Pseudocode for the Iterative Calibration Algorithm}

For completeness, we provide a pseudocode representation of the
iterative calibration procedure used in this work.

\begin{enumerate}
    \item Measure a representative intensity-response curve
    \(I=G(V_{AM})\).

    \item Numerically compute the inverse function
    \(g(I)=G^{-1}(I)\).

    \item Measure the initial frequency-dependent normalized response
    \(f_0(\omega)\).

    \item Set the iteration number \(n=0\).

    \item Repeat until the intensity fluctuation is below the desired threshold:
    \begin{enumerate}
        \item Compute
        \[
        C_n=\min_{\omega}
        \left[f_{n-1}(\omega)f_n(\omega)\right].
        \]

        \item Compute
        \[
        A_n(\omega)=
        \frac{C_n}{f_{n-1}(\omega)f_n(\omega)}.
        \]

        \item Compute
        \[
        V_{AM,n}^{\mathrm{res}}(\omega)
        =g\!\left(A_n(\omega)\right).
        \]

        \item Apply \(V_{AM,n}^{\mathrm{res}}(\omega)\) during the
        frequency sweep.

        \item Measure the updated response \(f_{n+1}(\omega)\).

        \item Increment \(n\leftarrow n+1\).
    \end{enumerate}

    \item Return the final corrected voltage waveform
    \(V_{AM}^{\mathrm{res}}(\omega)\).
\end{enumerate}

\end{minipage}

\end{document}